\documentclass[]{tMPH2e}

\usepackage{graphicx}

\usepackage{amssymb}

\usepackage{amsmath}

\usepackage[usenames]{color}

\def\etal{\mbox{et al.}}

\newcommand\beq{\begin{equation}}

\newcommand\eeq{\end{equation}}
\newcommand\beqa{\begin{eqnarray}}
\newcommand\eeqa{\end{eqnarray}}
\newcommand{\nn}{\nonumber\\}

\def\bal#1\eal{\begin{align}#1\end{align}}

\newcommand{\zero}{{(0)}}
\newcommand{\one}{{(1)}}
\newcommand{\two}{{(2)}}
\newcommand{\three}{{(3)}}

\newcommand{\la}{\lambda}

\begin{document}
\doi{10.1080/0026897YYxxxxxxxx}
 \issn{1362–3028}
\issnp{0026–8976}
\jvol{00}
\jnum{00} \jyear{2016} %\jmonth{10 May}

%\markboth{M. L\'opez de Haro,S. B. Yuste and A. Santos}{Molecular Physics}

\title{{\itshape Theoretical approaches to the structural properties of the square-shoulder fluid}}

\author{M. L\'opez de Haro$^{a}$$^{\ast}$\thanks{$^\ast$Email: malopez@unam.mx}, S. B. Yuste$^{b}$$^{\dagger}$\thanks{$^\dagger$Email: santos@unex.es} and
 A. Santos$^{b}$$^{\ddagger}$\thanks{$^\ddagger$Corresponding author. Email: andres@unex.es}
\\\vspace{6pt}  $^{a}${\em{Instituto de Energ\'{\i}as Renovables, Universidad Nacional Aut\'onoma de M\'exico (U.N.A.M.), Temixco, Morelos 62580, M{e}xico}};\\
%{\em{Departamento de F\'{\i}sica and Instituto de Computaci\'on Cient\'ifica Avanzada (ICCAEx), Universidad de
%Extremadura,  E-06071 Badajoz, Spain}};\\
$^{b}${\em{Departamento de F\'{\i}sica and Instituto de Computaci\'on Cient\'ifica Avanzada (ICCAEx), Universidad de
Extremadura,  E-06071 Badajoz, Spain}}}

\maketitle

\begin{abstract}
A comparison of simulation results with the prediction of the structural properties of square-shoulder fluids is carried out to assess the performance of three theories: Tang--Lu's first-order mean spherical approximation, the simplified exponential approximation of the latter and the rational-function approximation. These three theoretical developments share the characteristic of being analytical in Laplace space and of reducing in the proper limit to the Percus--Yevick result for  the hard-sphere fluid. Overall, the best agreement with the simulation data is obtained with the simplified exponential approximation.

 \bigskip
\begin{keywords}
radial distribution function;
square-shoulder potential;
first-order mean spherical approximation;
simplified exponential approximation;
rational-function approximation
\end{keywords}
\bigskip

\end{abstract}

\textit{This paper is dedicated {to the} memory of Dr.\ Yiping Tang}

\section{Introduction}

Simple models of the intermolecular potential {describing} a fluid are often useful to gain insight into many interesting phenomena occurring in real fluids. This is the case of the `square-shoulder' (SS) interaction, a purely repulsive potential, first used in this context by Hemmer and Stell \cite{HS70,SH72}, which has been the subject of many papers in the literature, including some rather recent ones \cite{SY76,YA77,LK93,LALR02,BF94,BF97,RVMN97,M98,GGJB99,LKLLW99,ZK01,MP03,RS03,PK08,BMAGPSSX09,BSB09,ZS09,FK10,GSC10,BL10,YSH11,HHT13,KH13}. This model may be considered to be the simplest one of the family of core-softened potentials that have been employed to study systems such as water \cite{BSB09}, metallic systems \cite{SY76}, colloidal suspensions \cite{LK93,LALR02} and aqueous solutions of electrolytes \cite{GGJB99}. The expression for the SS potential reads
\begin{equation}
\phi_{\text{SS}}(r)=\left\{
\begin{array}{ll}
\infty , & r<\sigma, \\
\epsilon, & \sigma<r<\lambda \sigma, \\
0, & r>\lambda \sigma,
\label{SS}
\end{array}
\right.
\end{equation}
where $r$ is the distance, $\sigma$ is the diameter of the hard core, $\epsilon>0$ is the shoulder height and $(\lambda - 1) \sigma$ is the shoulder width. Note that the thermodynamic properties of the SS fluid only depend on three dimensionless parameters, namely the packing fraction $\eta\equiv (\pi/6) \rho \sigma^3$ ($\rho$ being the number density), the reduced temperature $T^*=k_B T/\epsilon$ ($k_B$ and $T$ being the Boltzmann constant and the absolute temperature, respectively) and the width parameter $\lambda$. It is known in particular that the SS potential may lead to an isostructural solid--solid transition \cite{BF94,BF97}, to a fluid--solid coexisting line with a maximum melting temperature \cite{YA77},  to unusual phase behaviour \cite{RVMN97,M98,BMAGPSSX09} and to a rich variety of (self-organised) ordered structures \cite{ZK01,MP03,PK08,FK10}.

On the other hand, it is {interesting to note that} the SS potential becomes equivalent to a hard-sphere (HS) interaction of diameter $\sigma$ in the {limits of vanishing shoulder height ($\epsilon\to 0$) or  width}  ($\lambda\to 1$), and to an HS interaction of diameter $\la\sigma$ in the {limit of infinite shoulder height ($\epsilon\to \infty$).}
{These three limiting situations imply that}
\begin{subequations}
\label{limits}
\bal
\lim_{T^*\to\infty}g_{\text{SS}}(r;\sigma,\lambda,\eta,T^*)=&g_{\text{HS}}(r;\sigma,\eta),
\label{limit1}
\\
\lim_{\lambda\to 1}g_{\text{SS}}(r;\sigma,\lambda,\eta,T^*)=&g_{\text{HS}}(r;\sigma,\eta),
\label{limit2}
\\
\lim_{T^*\to 0}g_{\text{SS}}(r;\sigma,\lambda,\eta,T^*)=&g_{\text{HS}}(r;\lambda\sigma,\lambda^3\eta),
\label{limit3}
\eal
\end{subequations}
{where $g_{\text{SS}}(r;\sigma,\lambda,\eta,T^*)$ is the radial distribution function (RDF) of the SS fluid} {and $g_{\text{HS}}(r;\sigma,\eta)$ is the RDF of the HS fluid.} {Also, in the low-density limit one has} {$g_{\text{SS}}\to e^{-\phi_{\text{SS}}/k_BT}$, i.e.}
\begin{equation}
\lim_{\eta\to 0}g_{\text{SS}}(r;\sigma,\lambda,\eta,T^*)=\left\{
\begin{array}{ll}
0 , & r<\sigma, \\
e^{-1/T^*}, & \sigma<r<\lambda \sigma, \\
1, & r>\lambda \sigma.
\label{limit4}
\end{array}
\right.
\end{equation}
{Furthermore, continuity of} {$g_{\text{SS}}(r)\exp\left[\phi_{\text{SS}}(r)/T^*\right]$ at $r=\lambda\sigma$ implies the exact property}
\beq
g_{\text{SS}}(r=\lambda\sigma^+;\sigma,\lambda,\eta,T^*)=g_{\text{SS}}(r=\lambda\sigma^-;\sigma,\lambda,\eta,T^*)e^{1/T^*}.
\label{condcont}
\eeq

Irrespective of all the previous interesting findings, no exact results for the thermodynamic or structural properties of the SS fluid have been derived up to now. Moreover, not even the Percus--Yevick (PY) closure for the Ornstein--Zernike (OZ) integral equation for this system has led to analytical results. Therefore, the available data come from other approximate theories, from numerical solutions of the OZ equation with various closures and {from} simulation. Lang {\etal} \cite{LKLLW99} studied theoretically the SS fluid using the optimised random-phase approximation and the numerical solution of the OZ equation with the Rogers--Young  closure, and also performed Monte Carlo {(MC)} simulations. Zhou and Solana \cite{ZS09} also reported {MC} simulations for this system and theoretical results based on a bridge function approximation to close the OZ equation. Further simulation data for the SS fluid and a parametrisation of the direct correlation function which quantitatively agrees with the numerical solution of the OZ equation with the PY closure were presented by Guill\'en-Escamilla {\etal} \cite{GSC10}.

A few years ago, we carried out a theoretical study of the structural properties of this system \cite{YSH11} using the {rational-function} approximation (RFA) methodology that had been earlier employed successfully for {other} systems \cite{HYS08,S16}. More recently, Hlushak {\etal} \cite{HHT13} considered both the {Tang--Lu} first-order mean spherical approximation(FMSA) {\cite{TL93,TL94a,TL94b,T03}} and {its associated} simplified exponential approximation (SEXP{/FMSA}) {\cite{TL97}} for the {RDF} of the SS fluid and reported additional {MC} data for the system.
{These three approaches (FMSA, SEXP/FMSA and RFA)} {are formulated in terms of the Laplace transform}
\beq
\label{2.1}
G(s)=\int_0^\infty d r \,e^{-sr}r g(r)
\eeq
{of $r g(r)$, where, for simplicity, we have dropped the subscript SS and the arguments} {$(\sigma,\la,\eta,T^*)$. Moreover, the three of them reduce to the exact solution of the PY} {approximation of the HS fluid (of diameter $\sigma$) \cite{W63,W64,T63} in the limits $\lambda\to 1$ or} {$T^*\to\infty$ (see Equations \eqref{limit1} and \eqref{limit2}),} {although only the RFA reduces to such a solution} {(of diameter $\la \sigma$) in the limit $T^*\to 0$ (see Equation \eqref{limit3}).}
The aim of this paper is to perform a comparison of the results arising in the above three theoretical approximations with simulation data in order to asses the merits and limitations of each formulation.

The paper is organised as follows. In order to make it self-contained, in {Section \ref{sec2}} we present the main steps leading to derivation of the structural properties of the SS fluid using the three theoretical approaches referred to above, namely  the FMSA, the SEXP{/FMSA} and the RFA. This is followed in Section  \ref{sec3} by a comparison of the results of the different analytical approximations and those obtained from simulation. The paper is closed in Section  \ref{sec4} with further discussion and some concluding remarks.

\section{Radial distribution function of the square-shoulder fluid
\label{sec2}}

\subsection{{General properties}}
We begin by recalling general results that apply exactly to fluids whose molecules interact via \emph{any} intermolecular potential having a hard core at $r=\sigma$, as is the case of the SS fluid. {In what follows we will set, without loss of generality, $\sigma=1$. This means that} {henceforth all distances will be measured in units of the hard-core diameter $\sigma$.} In order to determine the structural properties of such fluids, it is convenient to deal with the Laplace transform  {defined in Equation \eqref{2.1}}.
Also, for later use, we introduce the auxiliary function $F(s)$ defined through \cite{YS91,HYS08,S16}
\beq
\label{2.2}
G(s)=\frac{sF(s)e^{-s}}{1+12\eta F(s)e^{-s}}.
\eeq

{Due to the hard-core condition} $g(r)=0$ for $r<1$, while $g(1^+)=\text{finite}$, one
has
\beq
\label{3.2s}
g(r)=\Theta(r-1)\left[g(1^+)+
g'(1^+)(r-1)+\cdots\right],
\eeq
where $g'(r)\equiv \partial g(r)/\partial r$ and $\Theta \left(x\right)$ is the Heaviside step function. In view of Equation \eqref{3.2s}, the large-$s$ {behaviours} of $G(s)$ {and $F(s)$ are} constrained, so that
\begin{subequations}
 \bal
\label{3.3s}
 e^{s}s G(s)=& g(1^+ )
+\left[g(1^+ )+ g'(1^+)\right] s^{-1}+{\cal
O}(s^{-2}),\\
F(s)=&g(1^+)s^{-2}+\left[g(1^+)+g'(1^+)\right]s^{-3}+\mathcal{O}(s^{-4}).
\eal
\end{subequations}
Therefore,
\beq
\label{2.3}
\lim_{s\to\infty} e^{s}sG(s)=\lim_{s\to\infty}s^{2}F(s)=
g(1^+)=\text{finite}.
\eeq

On the other hand,  the (reduced) isothermal compressibility is given by
\bal
\label{chi}
\chi \equiv& k_BT \left(\frac{\partial \rho}{\partial
p}\right)_T=1+24\eta\int_0^\infty d r\, r^2 \left[g(r)-1\right]\nn
=&1-24\eta\lim_{s\to 0}\frac{\partial}{\partial
s}\int_0^\infty d r\, e^{-s r}r\left[g(r)-1\right]\nn
=&1-24\eta\lim_{s\to 0}\frac{\partial}{\partial
s}\left[G(s)-s^{-2}\right].
\eal
Note that $\chi $ must also be
finite, and so  $\int_0^\infty d r\,r^2\left[g(r)-1\right]=\text{finite}$. Therefore, the \emph{weaker}
condition
 $\int_0^\infty d r \,r\left[g(r)-1\right]=\lim_{s\to 0}[G(s)-s^{-2}]=\text{finite}$
must hold. Taking {those} constraints into account leads to the following {small-$s$ behaviours:}
\begin{subequations}
   \bal
  \label{s2GG}
  s^2 G(s)=&1+\mathcal{O}(s^2),\\
  \label{2.4}
   F(s)=&-\frac{1}{12\eta}\left[1+s+\frac{s^2}{2}+\frac{1+2\eta}{12\eta}s^3+\frac{1+\eta/2}{12\eta}s^4
   \right]+\mathcal{O}(s^5).
  \eal
\end{subequations}

{The introduction of the auxiliary function $F(s)$ in Equation \eqref{2.2} allows us to obtain} {convenient expressions for the RDF in the coordination shells $n<r<n+1$. We first} rewrite Equation \eqref{2.2} as
\beq
    G(s)=\sum_{n=1}^\infty \left(-12\eta\right)^{n-1}s\left[F(s)\right]^n e^{-n s}.
    \label{10.14}
    \eeq
{Then,} the {RDF} may be obtained from
\beq
      g(r)=\frac{1}{r}\sum_{n=1}^\infty \left(-12\eta\right)^{n-1}{\psi}_{n}(r-n){\Theta(r-n)},
      \label{g(r)}
    \eeq
    where
{$\psi_n(r)$ is the inverse Laplace transform of $s\left[F(s)\right]^n$.}

Clearly, knowledge of {$F(s)$} immediately yields $G(s)$  and {$g(r)$. Irrespective} of how $G(s)$ is determined, once it is available another general result is that the static structure factor $S(q)$ (where $\mathbf{q}$ is the wavevector) of the fluid may be readily obtained from
\beq
\label{S(q)}
S(q)=1+\rho \int d\mathbf{r} \,e^{-\text{i} \mathbf{q}\cdot\mathbf{r}}[g(r)-1]= 1- 12 \eta \left.\frac{G(s)-G(-s)}{s}\right|_{s=\text{i}q},
\eeq
where $\text{i}$ is the imaginary unit. This confirms the important role played by $G(s)$ in the derivation of the structural properties of hard-core fluids.

\subsection{{Solution of the Percus--Yevick integral equation for hard-sphere fluids}}
{As is well known, the PY integral equation is exactly solvable for the HS fluid \cite{W63,W64,T63}.}
{In such a solution, the Laplace transform is given by Equation \eqref{2.2} with the auxiliary function} { $F(s)$ expressed as the simplest rational function complying with the physical}  {requirements \eqref{2.3} and \eqref{2.4} \cite{YS91,HYS08,S16}. More specifically,}
\beq
\label{A2}
G_0(s)=\frac{sF_0(s)e^{-s}}{1+12\eta F_0(s)e^{-s}},\quad F_0(s)=\frac{L_0(s)}{R_0(s)},
\eeq
where
\begin{subequations}
\label{A3+4}
\bal
L_0(s)=&1+2\eta+(1+\eta/2)s,
\label{A3}
\\
R_0(s)=&-12\eta(1+2\eta)+18\eta^2 s+6\eta(1-\eta)s^2+(1-\eta)^2s^3.
\label{A4}
\eal
\end{subequations}
{The subscript $0$ in Equation \eqref{A2} means that the solution is restricted to HS systems} {with a diameter $\sigma=1$.}
{In the case of HS systems with a diameter $\lambda\sigma=\lambda$,} {the PY solution is given by Equations \eqref{A2} and \eqref{A3+4} with the replacements
$G_0\to \lambda^{-2}G_0$,} {$s\to \lambda s$, and $\eta\to\lambda^3 \eta$ \cite{YSH11}. In real space, according to Equation \eqref{g(r)}, the HS RDF is}
{given by \cite{LNP_book_note_13_10}}
\beq
      g_0(r)=\frac{1}{r}\sum_{n=1}^\infty \left(-12\eta\right)^{n-1}{\psi}_{0n}(r-n){\Theta(r-n)},
      \label{g_0(r)}
    \eeq
{with}
\begin{subequations}
\label{10.X1}
 \bal
\psi_{0n}(r)=&\sum_{j=1}^n \frac{\sum_{i=1}^3 a_{n j}^{(i)}e^{s_{0i} r}}{(n-j)!(j-1)!}r^{n-j}\;,
\label{10.X1.a}\\
a_{n j}^{(i)}=&\lim_{s\to s_{0i}}\left(\frac{\partial}{\partial s}\right)^{j-1}\left\{s\left[(s-s_{0i})F_0(s)\right]^n\right\},
\eal
\end{subequations}
{where $\{s_{0i}, i=1,2,3\}$ are the three roots  of the cubic equation $R_0(s)=0$.}

\subsection{{Tang and Lu's FMSA and SEXP/FMSA  for the structural properties} {of the square-shoulder fluid}\label{appA}}

In this subsection, we consider the results for the structural properties of the SS fluid as derived from the FMSA and SEXP/FMSA developed by Tang and Lu \cite{TL94a,TL94b,TL97}. The {presentation}  follows very closely the one {given} for the square-well (SW) fluid in Ref.\ \cite{LSYS05} with the simple replacement of $T^*$ by $-T^*$.

\subsubsection{{FMSA}}
The {FMSA} theoretical approach, {applicable to}  potentials with a spherical hard core and an arbitrary tail, consists {in} combining thermodynamic perturbation theory (taking  the HS {system} as the reference fluid) and the mean spherical approximation (MSA) to derive an analytical solution to the OZ equation as a series {in powers of the inverse temperature $1/T^*$}. In the SS fluid case, the expansion of the {RDF} $g(r)$  to first order reads
\begin{subequations}
\bal
G(s) =& G_{0}(s) + G_{1}(s) \frac{1}{T^{*}},
\label{eq:G(s))-per}
\\
g(r) =& g_{0}(r) + g_{1}(r) \frac{1}{T^{*}},
\label{eq:g(x)-per}
\eal
\end{subequations}
{where the reference HS functions $G_0(s)$ and $g_0(r)$ are given by Equations \eqref{A2} and \eqref{g_0(r)},} {respectively.}
The first-order term $G_1(s)$ {is the opposite of the one} for the SW fluid \cite{TL94a,TL94b}, namely
\bal
G_1(s)=&\frac{(1-\eta)^4}{\left[Q_0(s)\right]^2}\left\{\frac{s^4(1+\lambda
s)}{\left[R_0(-s)\right]^2}e^{-\lambda s}   -
\sum_{i=1}^3\frac{s_{0i}^3e^{(\lambda-1)s_{0i}}}{(s+s_{0i})\left[R_0'(s_{0i})\right]^2}e^{-s} \right.\nn
&\left.\times
\left[\frac{s_{0i}(1-\lambda
s_{0i})}{s+s_{0i}}
+s_{0i}(1-\lambda s_{0i})\frac{R_0''(s_{0i})}{R_0'(s_{0i})}
-4+(1+4\lambda)s_{0i}+\lambda(\lambda-1)s_{0i}^2\right]
\right\},
\label{A5}
\eal
{where the primes in $R_0'(s)$ and $R_0''(s)$ denote} derivatives with respect to $s$, and
\beq
Q_0(s)\equiv\frac{R_0(s)+12\eta L_0(s)e^{-s}}{(1-\eta)^2s^3}.
\label{A6}
\eeq
{As in Equation \eqref{10.X1.a}, the summation over $i$}  in Equation (\ref{A5}) extends over the three zeros of
$R_0(s)$.
{It is straightforward to find from Equation \eqref{A5} the jump discontinuity} {of $g_1(r)$ at $r=\la$:}
\beq
g_1(\la^+)-g_1(\la^-)=\lim_{s\to\infty}\frac{s}{\la}\frac{(1-\eta)^4}{\left[Q_0(s)\right]^2}\frac{s^4(1+\lambda
s)}{\left[R_0(-s)\right]^2}=1.
\label{disc}
\eeq
{In what concerns the low-density limit $\eta\to 0$, one has}
\begin{subequations}
\bal
\lim_{\eta\to 0}G_1(s)=&s^{-2}(1+\la s)e^{-\la s}-s^{-2}(1+s)e^{-s},\\
\lim_{\eta\to 0}g_1(r)=&\Theta(r-\la)-\Theta(r-1).
\label{lim_eta}
\eal
\end{subequations}

Performing the inverse Laplace transform of $G_1(s)$ one can get explicit expressions
for $g_1(r)$ inside the shells $n< r < n+1$, which become
increasingly more complicated as $n$ grows. Due to the fact that they are not very illuminating and may be found elsewhere \cite{HHT13}, they will be  omitted {here. Alternatively, $G_1(s)$ can be numerically inverted \cite{AW92} to obtain $g_1(r)$}.

\subsubsection{{SEXP/FMSA}}

Now we turn to the SEXP{/FMSA \cite{TL97}}. In this theory, the {RDF} of the SS fluid is approximated by
\begin{equation}
g(r)=g_{0}(r)\exp\left[\frac{g_{1}(r)}{T^*}\right],
\label{gexp}
\end{equation}
where $g_{0}(r)$ and $g_{1}(r)$ are those of the FMSA theory. This simplified exponential approximation ensures the positive definite character of the {RDF.} Notice that the expansion of the SEXP/FMSA \eqref{gexp} to first order in $1/T^*$ differs from the FMSA  \eqref{eq:g(x)-per}.

{In the high-temperature limit $T^*\to\infty$, $g(r)\to g_0(r)$ both in the FMSA and in} {the SEXP/FMSA, so that Equation \eqref{limit1} is verified. A less trivial consistency test} {corresponds to the limit $\lambda\to 1$.} {In that case, taking into account the identity}
\beq
\frac{s^4(1+s)}{\left[R_0(-s)\right]^2}=
\sum_{i=1}^3\frac{s_{0i}^3}{(s+s_{0i})\left[R_0'(s_{0i})\right]^2}\left[\frac{s_{0i}(1-s_{0i})}{s+s_{0i}}+s_{0i}(1- s_{0i})\frac{R_0''(s_{0i})}{R_0'(s_{0i})}
-4+5s_{0i}\right],
\eeq
{one finds that $\lim_{\lambda\to 1}G_1(s)=0$, so that $g(r)\to g_0(r)$ in that limit, in agreement} {with Equation \eqref{limit2}.
As for conditions \eqref{limit4} and \eqref{condcont}, Equations \eqref{disc} and \eqref{lim_eta} show that they are satisfied by the SEXP/FMSA} {at finite $T^*$, but not by the FMSA (except to first order in $1/T^*$).}

{On the other hand, in the zero-temperature limit $T^*\to 0$ both the FMSA \eqref{eq:g(x)-per}} {and the SEXP/FMSA \eqref{gexp} become singular and none of them complies with Equation \eqref{limit3}.}

\subsection{The {rational-function} approximation\label{RFAAA}}

{We now} outline the main steps of the RFA approach to the structural properties of the SS fluid. For further details the reader is referred to Ref.\ \cite{YSH11}.

In the RFA it is assumed that the auxiliary function {$F(s)$ in Equation \eqref{2.2}} takes the following form
 \beq
\label{eq:F(t)}
 F(s)=\frac{L(s)}{R(s)},
 \eeq
where
\begin{subequations}
\bal
L(s)=&1+2\eta-K^\zero+L^\one s+e^{-(\la-1)}\left[K^\zero+K^\one s\right],
\\
R(s)=&-12\eta(1+2\eta)+R^\one s+R^\two s^2+R^\three s^3.
\eal
\end{subequations}
{Enforcement of Equation \eqref{2.4} allows one to express the coefficients $L^\one$, $R^\one$, $R^\two$ and $R^\three$} as linear functions of {$K^\zero$ and $K^\one$:}
\begin{subequations}
\label{c5-c7}
 \bal
\label{c5}
L^\one=&1+\frac{\eta}{2}+(\lambda-1)\frac{2+\eta\left(\la^3+\la^2+\la+1\right)}
{2(1+2\eta)} {K^\zero}-\frac{1+2\lambda^3\eta}{1+2\eta}
{K^\one},
\\
\label{c6}
R^\one=&{6\eta^2}\left[{3}-
(\lambda-1)^2\frac{\lambda^2+2\lambda+3}{1+2\eta}{K^\zero}+\frac{4(\lambda^3-1)}{1+2\eta} {K^\one}\right],
\\
\label{c7}
R^\two=&{6\eta}\left[1-\eta-(\lambda-1)^2\frac{1-\eta(\lambda+1)^2}{1+2\eta}K^\zero+2(\lambda-1)\frac{1-2\eta
\lambda(\lambda+1)}{1+2\eta}K^\one\right],
\\
\label{c7bis}
R^\three=&(1-\eta)^2+\eta(
\lambda-1)^2\frac{4+2\lambda-\eta(3\lambda^2+2 \lambda+1)}{1+2\eta}K^\zero
\nn &
-6\eta(\lambda-1)
\frac{\lambda+1-2\eta \lambda^2}{1+2\eta}K^\one,
\eal
\end{subequations}
{Next, application of condition \eqref{condcont} gives \cite{YSH11,HYS08,S16}}
\beq
\sum_{i=1}^3 \frac{1+2\eta-K^\zero+K^\one s_i}{R'(s_i)}s_i
e^{(\la-1)s_i}=\frac{K^\one}{\left(e^{1/T^*}-1\right)R^\three},
\label{SWB7}
\eeq
{where $\{s_{i}, i=1,2,3\}$ are the three roots  of the cubic equation $R(s)=0$ and} {it has been assumed that $\la\leq 2$.}
To close the description,
the coefficient {$K^\zero/(1+2\eta)$}  is fixed at its exact zero-density limit value, namely \cite{YSH11}
\beq
K^\zero=(1+2\eta)(1-e^{-1/T^*}) .
\label{Q0}
\eeq
Therefore, Equation \eqref{SWB7} becomes
a {transcendental} equation for {$K^\one$} that needs to be solved numerically.

Once the coefficients  {$L^\one$, $R^\one$, $R^\two$, $R^\three$, $K^\zero$ and $K^\one$ are} determined as functions of $\eta$, $T^*$ and $\lambda$ through Equations \eqref{c5-c7}--\eqref{Q0},  $G(s)$ becomes completely specified in the RFA. The {RDF} $g(r)$ in this case may again be obtained by {numerically} taking the inverse Laplace transform \cite{AW92} of the corresponding $G(s)$ or, equivalently, from the use of Equations \eqref{eq:F(t)}--\eqref{Q0}, together with Equation \eqref{g(r)} \cite{S16}.

{It can be easily  checked that the RFA is consistent with the exact conditions \eqref{limits}--\eqref{condcont} \cite{YSH11}.} {Table \ref{tablecond} summarises which ones of those conditions are fulfilled by the FMSA,} {SEXP/FMSA and RFA approaches.}

\begin{table*}
  \tbl{{Summary showing which ones of the exact conditions \protect\eqref{limits}--\protect\eqref{condcont} are fulfilled by the FMSA,} {SEXP/FMSA and RFA approaches.}}
{\begin{tabular}{@{}lccccc}\toprule
   Approximation& Equation \protect\eqref{limit1} & Equation \protect\eqref{limit2}& Equation \protect\eqref{limit3}& Equation \protect\eqref{limit4}& Equation \protect\eqref{condcont}\\
\colrule
FMSA&Yes&Yes&No&No&No\\
SEXP/FMSA&Yes&Yes&No&Yes&Yes\\
RFA&Yes&Yes&Yes&Yes&Yes\\
   \botrule
  \end{tabular}}
  \label{tablecond}
\end{table*}

\section{Comparison with simulation data \label{sec3}}

\begin{figure}
\begin{center}
\includegraphics[width=.6\columnwidth]{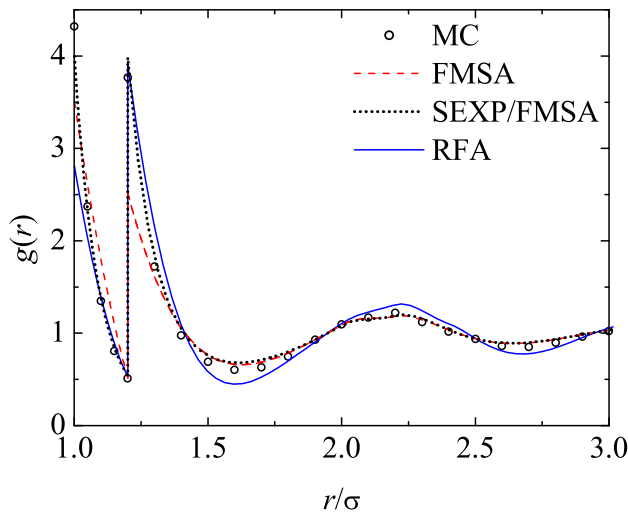}
\caption{Radial distribution function $g(r)$ as a function of distance
$r$ for an SS fluid having $\lambda=1.2$, $T^*=0.5$ and  $\eta=0.4$ ($\rho\sigma^3=0.764$)  as obtained from the FMSA (dashed line), the SEXP/FMSA (dotted line), the RFA (solid line) and simulation data from  Ref.\ \protect\cite{LKLLW99} (circles).\label{fig1}}
\end{center}
\end{figure}

\begin{figure}\begin{center}
\includegraphics[width=.6\columnwidth]{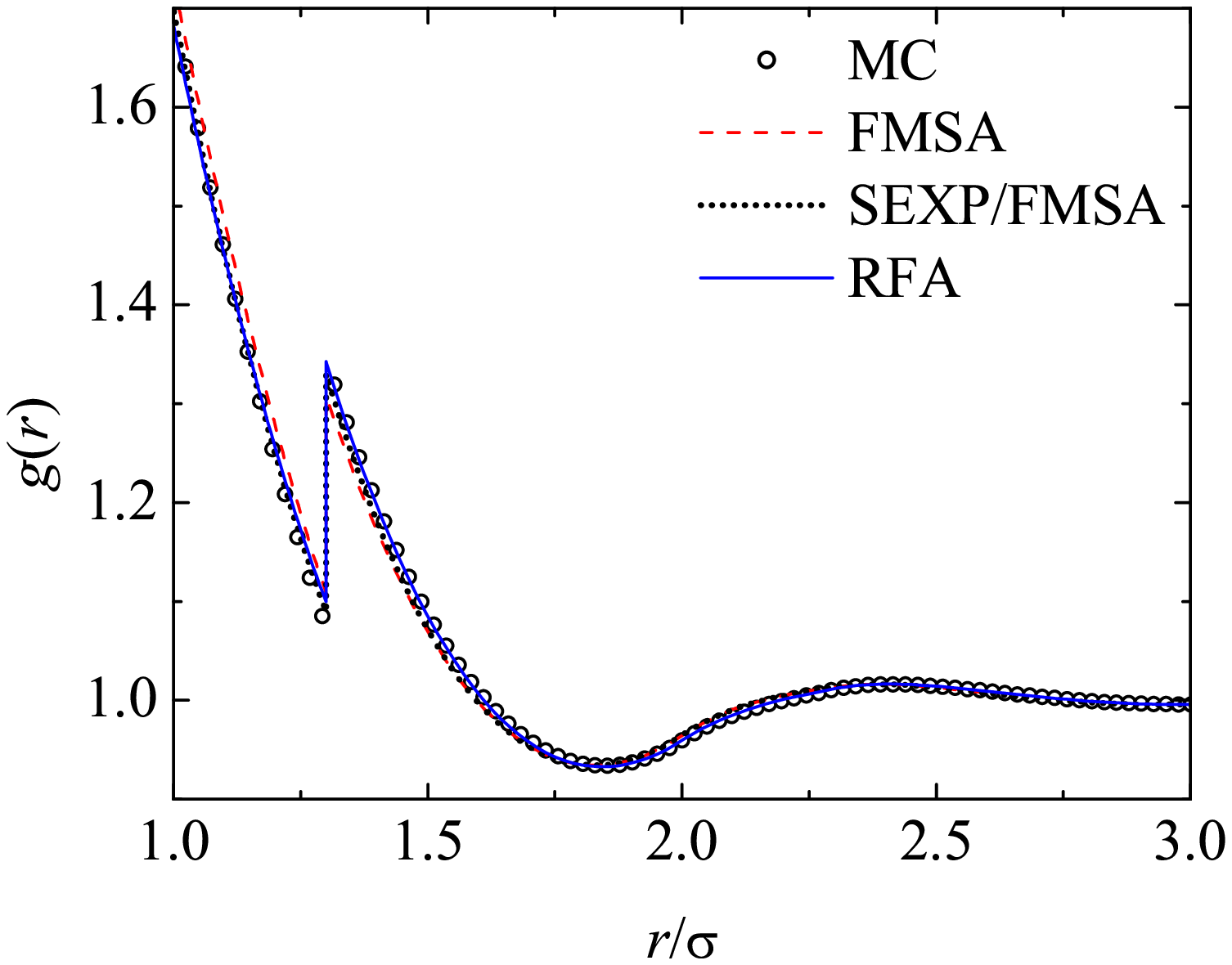}
\caption{Radial distribution function $g(r)$ as a function of distance
$r$ for an SS fluid having $\lambda=1.3$, $T^*=5$ and  $\eta=0.2094$ ($\rho\sigma^3=0.4$)  as obtained from the FMSA (dashed line), the SEXP/FMSA (dotted line), the RFA (solid line) and simulation data from  Ref.\ \protect\cite{GSC10} (circles).}
\label{fig2}
\end{center}
\end{figure}

\begin{figure}\begin{center}
\includegraphics[width=.6\columnwidth]{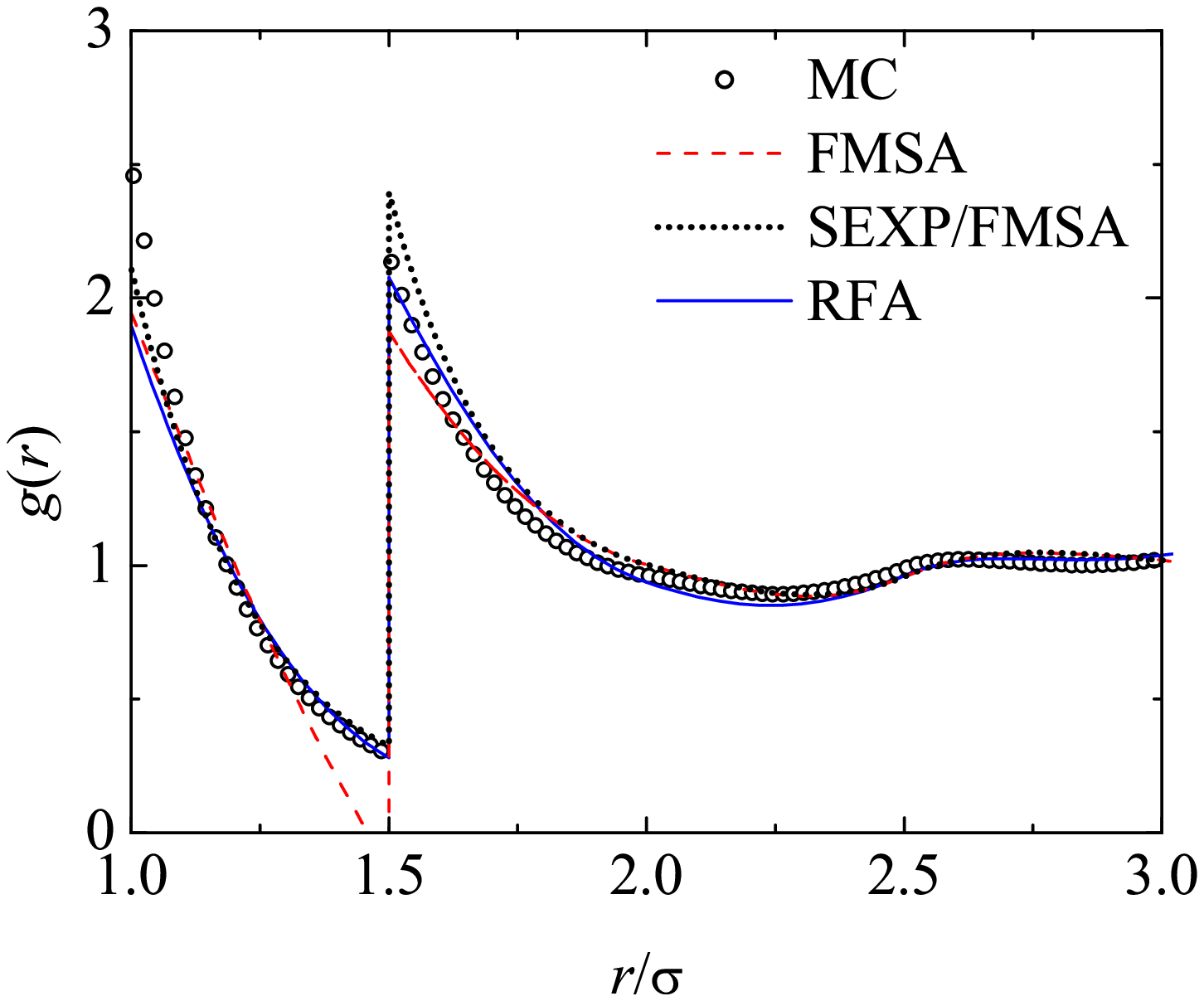}
\caption{Radial distribution function $g(r)$ as a function of distance
$r$ for an SS fluid having $\lambda=1.5$, $T^*=0.5$ and $\eta=0.2094$ ($\rho\sigma^3=0.4$) as obtained from the FMSA (dashed line), the SEXP/FMSA (dotted line), the RFA (solid line) and simulation data from  Ref.\ \protect\cite{ZS09} (circles).}
\label{fig3}
\end{center}
\end{figure}

\begin{figure}
\begin{center}
\includegraphics[width=.6\columnwidth]{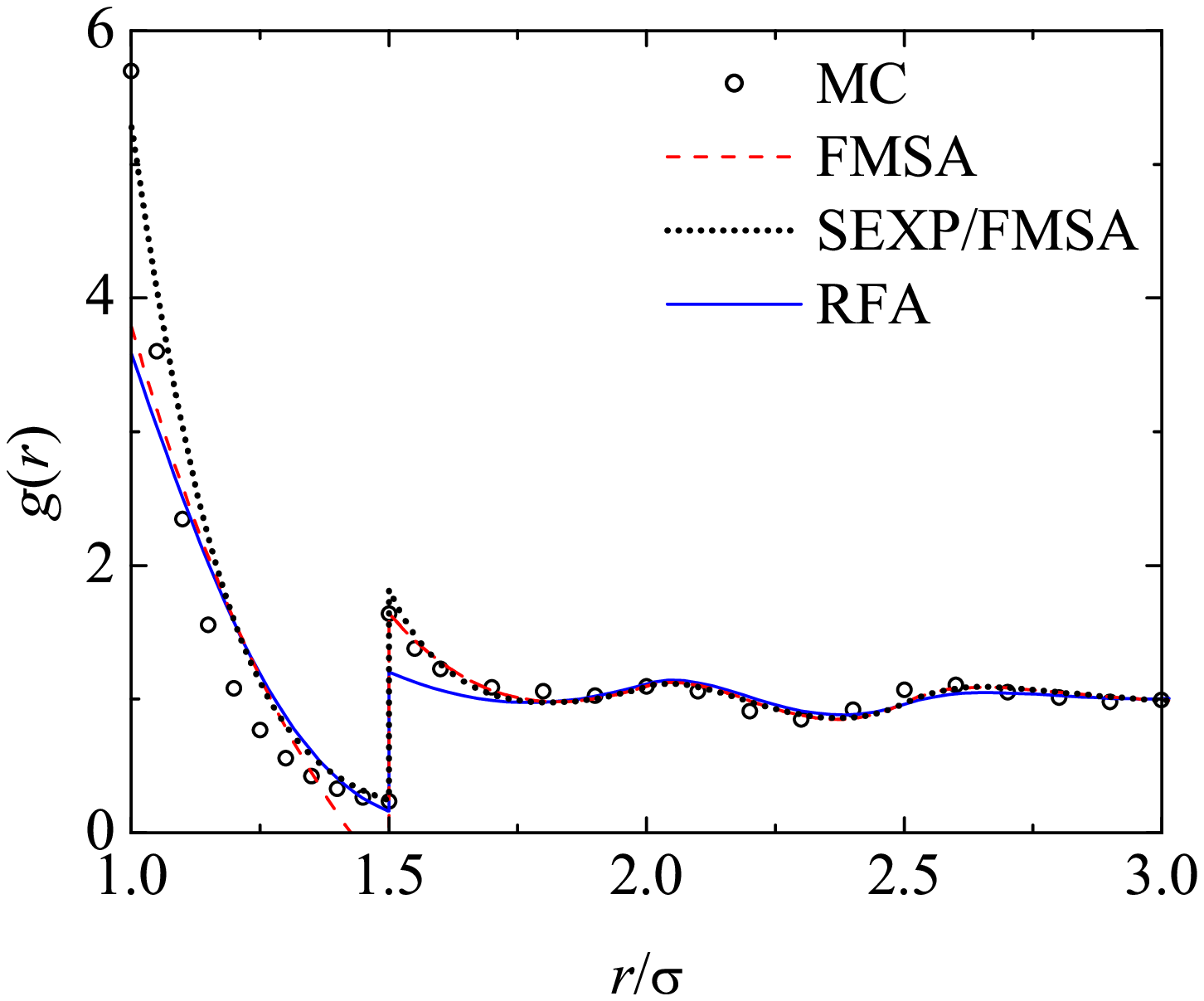}
\caption{Radial distribution function $g(r)$ as a function of distance
$r$ for an SS fluid having $\lambda=1.5$, $T^*=0.5$ and  $\eta=0.4$ ($\rho\sigma^3=0.764$)  as obtained from the FMSA (dashed line), the SEXP/FMSA (dotted line), the RFA (solid line) and simulation data from  Ref.\ \protect\cite{LKLLW99} (circles).\label{fig4}}
\end{center}
\end{figure}

\begin{figure}\begin{center}
\includegraphics[width=.6\columnwidth]{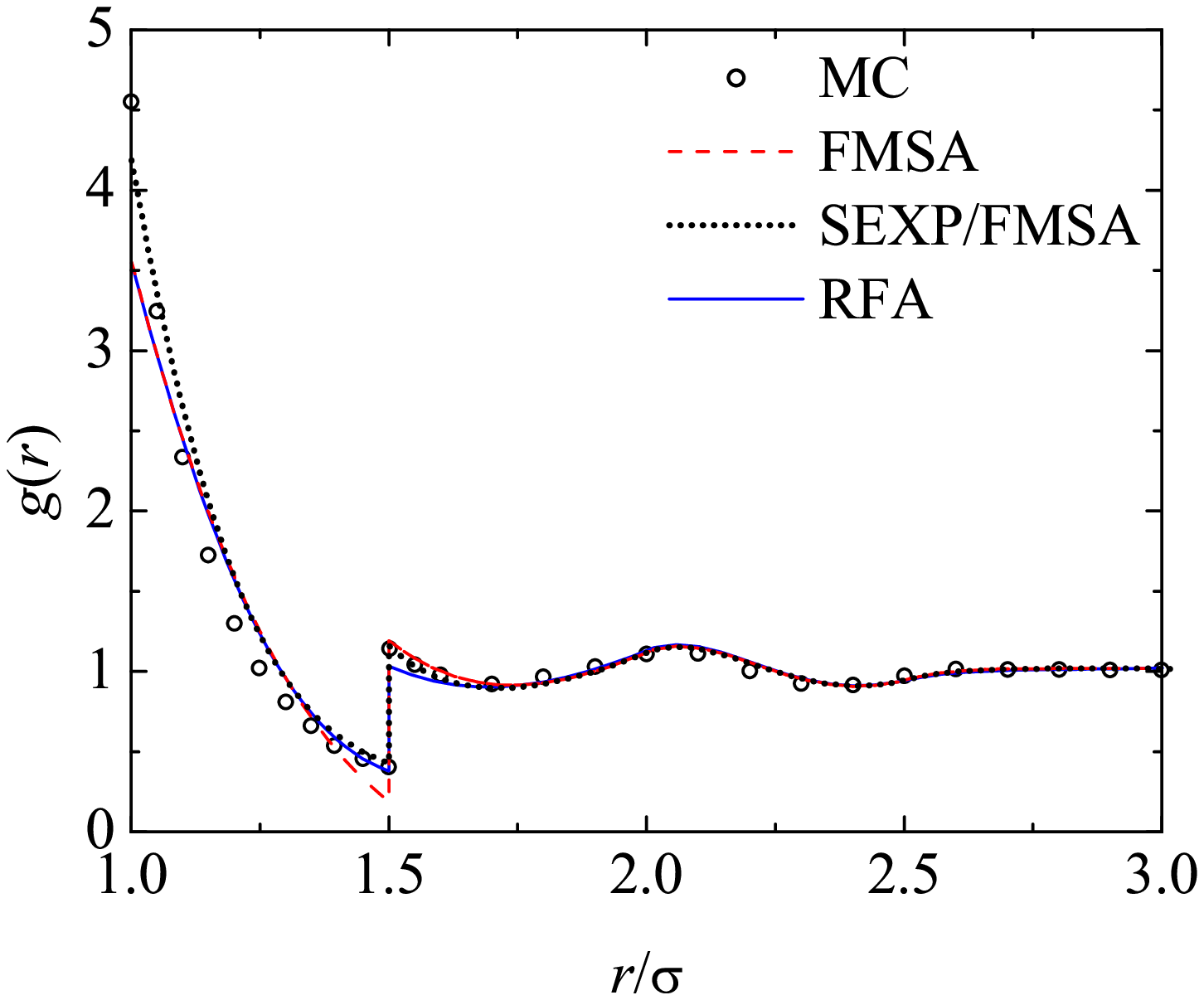}
\caption{Radial distribution function $g(r)$ as a function of distance
$r$ for an SS fluid having $\lambda=1.5$, $T^*=1$ and  $\eta=0.4$ ($\rho\sigma^3=0.764$)  as obtained from the FMSA (dashed line), the SEXP/FMSA (dotted line), the RFA (solid line) and simulation data from  Ref.\ \protect\cite{LKLLW99} (circles).}
\label{fig5}
\end{center}
\end{figure}

\begin{figure}\begin{center}
\includegraphics[width=.6\columnwidth]{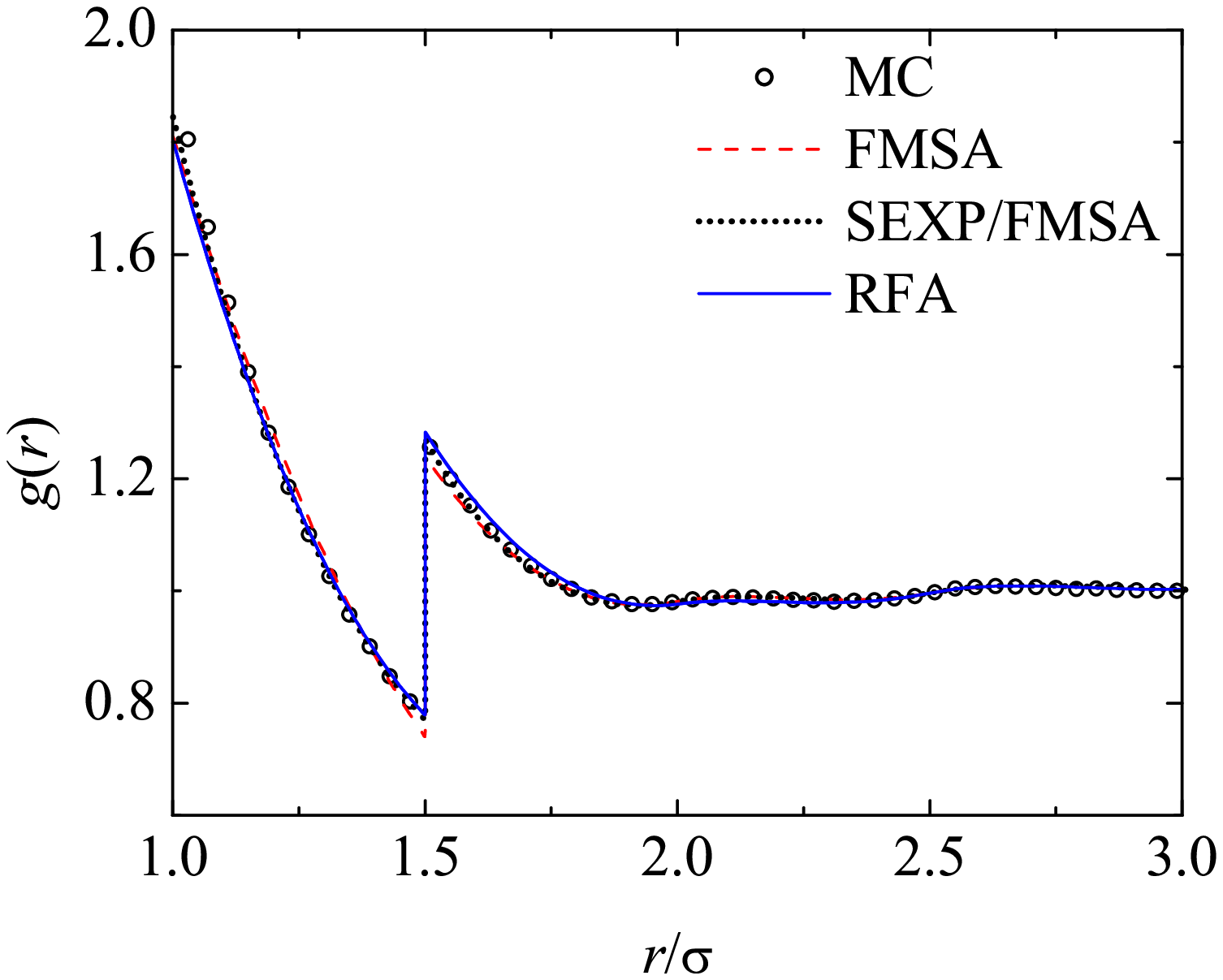}
\caption{Radial distribution function $g(r)$ as a function of distance
$r$ for an SS fluid having $\lambda=1.5$, $T^*=2$ and  $\eta=0.2094$ ($\rho\sigma^3=0.4$)  as obtained from the FMSA (dashed line), the SEXP/FMSA (dotted line), the RFA (solid line) and simulation data from Ref.\ \protect\cite{GSC10} (circles).}
\label{fig6}
\end{center}\end{figure}

\begin{figure}\begin{center}
\includegraphics[width=.6\columnwidth]{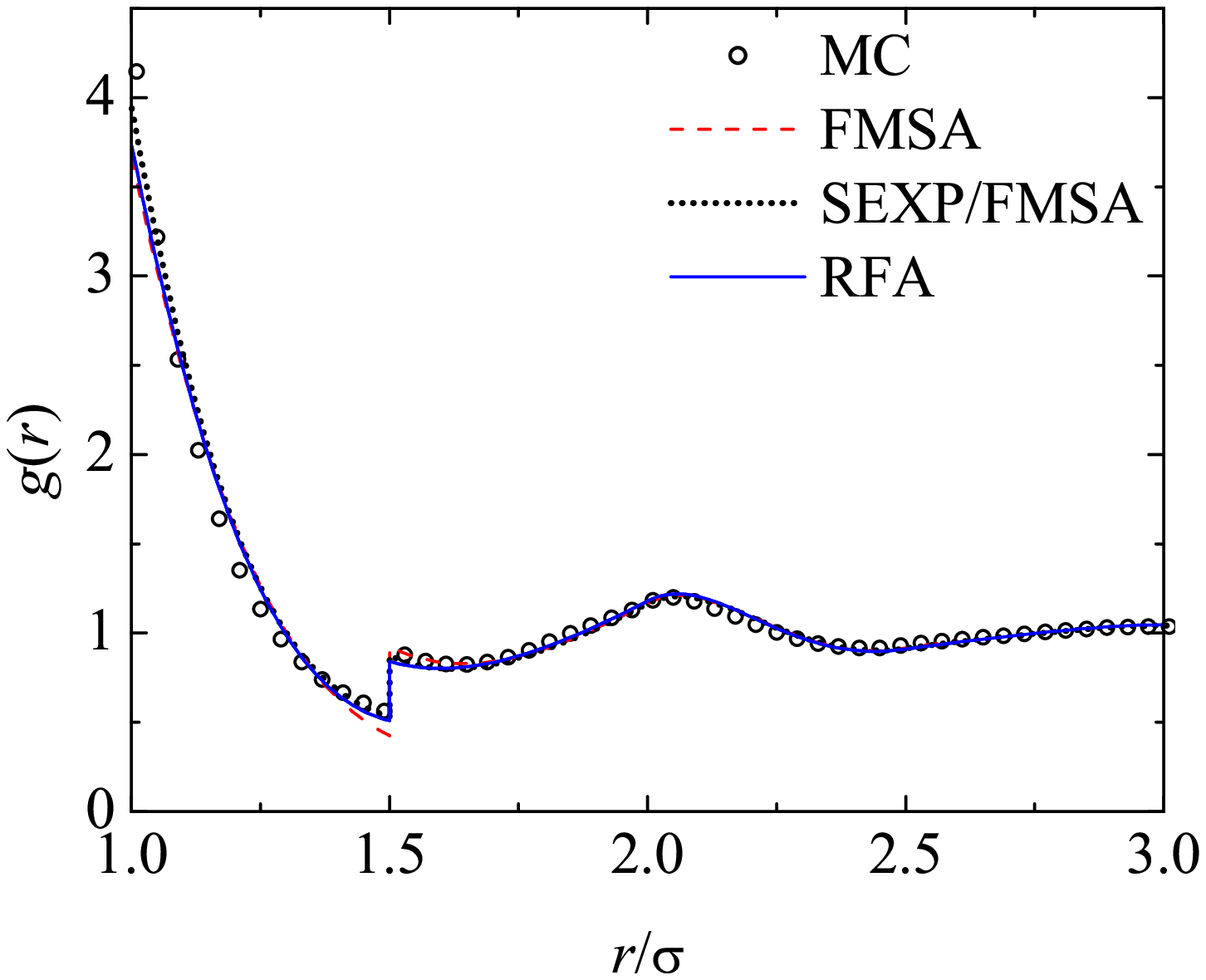}
\caption{Radial distribution function $g(r)$ as a function of distance
$r$ for an SS fluid having $\lambda=1.5$, $T^*=2$ and  $\eta=0.4189$ ($\rho\sigma^3=0.8$)  as obtained from the FMSA (dashed line), the SEXP/FMSA (dotted line), the RFA (solid line) and simulation data from Ref.\ \protect\cite{GSC10} (circles).}
\label{fig7}
\end{center}\end{figure}

In order to assess the value of the three previous theoretical approximations for the structural properties of SS fluids, in this section we carry out a comparison between the results derived from them and those obtained from simulation \cite{LKLLW99,ZS09,GSC10}. Although we have made such a comparison with many other data, in Figures \ref{fig1}--\ref{fig7} we only show graphs of $g(r)$ vs $r$ for some representative cases.

Figure \ref{fig1} displays the {RDF}s computed with the three theories and the corresponding simulation data for a rather narrow shoulder ($\la=1.2$) at relatively low temperature ($T^*=0.5$) and  high density ($\eta=0.4\Rightarrow \lambda^3\eta=0.6912$). It is clear that none of the theories provides a full quantitative agreement with simulation. In particular, the contact value $g(1^+)$ is underestimated by all of them, with the SEXP/FMSA giving the closest estimate. While $g(\lambda^-)$ is well accounted for in all instances, in the case of $g(\lambda^+)$ both the RFA and the SEXP/FMSA do a reasonable job but the FMSA heavily underestimates its value. This is obviously related to the failure of the FMSA to account for the exact relation \eqref{condcont}. The location of the maxima and minima of the subsequent oscillations seem to be well captured by all approaches, the RFA exhibiting the poorest agreement with the actual values of the function.

As a representative example of a wider shoulder, Figure \ref{fig2} presents the case $\lambda=1.3$ at  high temperature ($T^*=5$) and medium density ($\eta=0.2094$). In this case, the three theories attain very good quantitative agreement, even for the contact value $g(1^+)$ and the shoulder-edge values $g(\lambda^-)$ and $g(\lambda^+)$. At $T^*=5$, $e^{1/T^*}=1.221\simeq 1+ 1/T^*$ and thus condition \eqref{condcont} can be replaced by its linearised version, which is satisfied by the FMSA. Although not shown, we have checked that the good agreement observed in Figure \ref{fig2} stays rather reasonable as one increases the packing fraction up to its double value, provided the values of $T^*=5$ and $\lambda=1.3$ are maintained.

Next, in Figures \ref{fig3}--\ref{fig7} we fix the standard value $\la=1.5$ and analyse the influence of density and, especially, temperature. When  $T^*=0.5$ and $\eta=0.2094$ (see Figure \ref{fig3}), the situation is somewhat similar to the case in Figure \ref{fig1}, with the contact value being underestimated by all the theories but much less than in the denser system considered in Figure \ref{fig1}. This time, although the quantitative agreement seems not to be  bad for both the RFA and the SEXP/FMSA, none of the theories is completely satisfactory. In this instance, it is the RFA the one that does the best job, while the FMSA fails badly, especially in the prediction of both $g(\lambda^-)$ (that is even negative) and $g(\lambda^+)$. As the density is increased to $\eta=0.4$  (see Figure \ref{fig4}), the region between contact and $r=\lambda^-$ is poorly described by all theories. Further, the RFA underestimates heavily $g(1^+)$, while both the FMSA and the SEXP/FMSA seem to do a reasonable job for $r>\lambda^+$.

Figure \ref{fig5} corresponds to the same density as in Figure \ref{fig4}, but the temperature has been doubled. Now, as expected, the situation improves for the FMSA. Moreover, the SEXP/FMSA does the best performance.

We keep increasing temperature to $T^*=2$ in Figures  \ref{fig6} (corresponding to $\eta=0.2094$) and \ref{fig7} (corresponding to $\eta=0.4189$). In the former case, the agreement between theories and simulation is very good, with an almost equal performance of the RFA and the SEXP/FMSA, which are both only slightly superior to the FMSA. Comparison between Figures \ref{fig3} and \ref{fig6} shows the significant improvement of the FMSA as temperature increases from $T^*=0.5$ to $T^*=2$. Finally,  Figure \ref{fig7} shows that the trends observed in Figure \ref{fig6} still hold as one increases the packing fraction, although now the predictions for the contact value somewhat worsen.

It is clear that the poorest performance of all the theoretical developments occurs for relatively low reduced temperatures, as the cases with $T^*=0.5$ (see Figures \ref{fig3} and \ref{fig4}) illustrate. For such temperatures  the deficiencies show mostly around the contact region but the theories become less reliable even beyond that region as the packing fraction is increased. {The influence of temperature at  fixed values of $\la$ and $\eta$ may be assessed by comparing Figures \ref{fig3} and \ref{fig6}. We note that, as the temperature decreases, the contact value increases moderately and also $g(r)$  for $r\gtrsim 2$ becomes more structured. The strongest influence of temperature occurs in the region  $r\approx\la$ and, as expected on physical grounds, the discontinuity  at $r=\la$ becomes much more pronounced as the temperature decreases.}

{Some insight into the effect of the shoulder width on the performance of the theories may be gained by comparing the cases with $T^*=0.5$ and  $\eta=0.4$ in Figures \ref{fig1} ($\la=1.2$) and \ref{fig4} ($\la=1.5$).  One may observe that shrinking the shoulder at fixed temperature and packing fraction makes the theories in general become more reliable, even in a case where the low value of the reduced temperature is less favourable for them. This is not surprising in view of the fact that the SS model becomes closer and closer to the HS model as the shoulder width decreases, as shown by Equation \eqref{limit2}. In this HS limit all three theories reduce to Wertheim--Thiele's \cite{W63,W64,T63} exact solution of the OZ equation with the PY closure. Since the HS potential is also reached from  the SS one in the high-temperature limit ($T^*\to\infty$) (see Equation \eqref{limit2}), a better performance of  all three approaches  can be expected to hold for sufficiently high temperatures. This is confirmed by Figure \ref{fig2} in the case $\la=1.3$, $T^*=5$ and $\eta=0.2094$. On the other hand, if the temperature were so low that the situation described by Equation \eqref{limit3} were approached, the RFA is expected to prevail over the FMSA and SEXP/FMSA.}

Before closing this section, it is worth noting that the systems considered in Figures \ref{fig1}, \ref{fig3} and \ref{fig4} are the same as those considered in the sixth panel of Figure 2, the third panel of Figure 3 and the sixth panel of Figure 3, respectively, of Ref.\ \cite{HHT13}. On the other hand, the curves corresponding to the SEXP/FMSA displayed in those figures of Ref.\ \cite{HHT13} exhibit a higher discrepancy with respect to simulation data than in our Figures  \ref{fig1}, \ref{fig3} and \ref{fig4}.
This is due to the fact that, through an involuntary mistake in  Ref.\ \cite{HHT13},  the SEXP/FMSA recipe \eqref{gexp} was actually applied only if $g_1(r)<0$, while it was replaced by its linearised version  $g(r)=g_0(r)[1+g_1(r)/T^*]$ if $g_1(r)>0$ \cite{private}.

\section{Concluding remarks}
\label{sec4}

In this paper we have revisited three analytical procedures \cite{TL94a,TL94b,TL97,HHT13,YSH11} to obtain the structural properties of SS fluids and compared their predictions against simulation results \cite{LKLLW99,ZS09,GSC10} in order to assess their merits and limitations. All these approaches have in common the fact that they are analytical in Laplace space and  reduce in two independent limits [see Equations \eqref{limit1} and \eqref{limit2}] to the PY result for  the HS fluid, although only the RFA does it in a third limit [see Equation \eqref{limit3}]. One should insist on the usefulness of having at hand analytical or semi-analytical approximations for the equilibrium structural properties of simple fluids. In this sense, the FMSA, the SEXP/FMSA and the RFA have once more proved their importance and are simple enough to allow for immediate computations. In a way, the results of the present paper are complementary to those of Refs.\ \cite{YSH11} and \cite{HHT13}, where a (partial) similar analysis was carried out.  We have confirmed that the theories lead to reasonably accurate results at any fluid packing fraction if the shoulder is sufficiently narrow (say $\la \leq 1.2$), as well as for any width if $\eta$ is small enough ($\eta \leq 0.4$). However, as the width and/or the packing fraction increase, the predictions  worsen, especially at low temperatures and in the region between contact and $\la$.
In any case, from our analysis we can conclude the following. Being a perturbation theory, as expected the FMSA works reasonably well at high temperatures, but worsens as the temperature is reduced, even yielding negative values for $g(\lambda^-)$, especially when $\lambda$ is increased. The RFA is a reasonable compromise between accuracy and simplicity, but presents some limitations when either the shoulder width or the packing fraction increase. Finally, we found that the best overall performance was shown by the SEXP/FMSA  and that it was even better than what was reported in Ref.\ \cite{HHT13}.

\section*{Acknowledgements}
Mariano L\'opez de Haro wants to thank the hospitality of Universidad de Extremadura (Spain), where the first {stages of this work were carried out}.
The authors are grateful to Stepan P. Hlushak for clarifying the reason for the discrepancy between our results and those of Ref. \cite{HHT13}  for the SEXP/FMSA approximation.

\section*{Disclosure statement}
No potential conflict of interest was reported by the authors.

\section*{Funding}
The research of Santos Bravo Yuste and Andr\'es Santos has been partially financed by the Ministerio de Econom\'ia y Competitividad (Spain) [grant number FIS2013-42840-P], the Regional Government of Extremadura (Spain) [grant number GR15104] (partially financed by the ERDF).

%\bibliographystyle{tMPH}
%\bibliography{D:/Dropbox/Public/bib_files/liquid}

\begin{thebibliography}{39}
\providecommand{\url}[1]{\texttt{#1}}
\providecommand{\urlprefix}{URL }
\markboth{Taylor \& Francis and I.T. Consultant}{Molecular Physics}

\bibitem{HS70}
P.C. Hemmer and G. Stell,  Phys. Rev. Lett.  \textbf{24}, 1284 (1970).

\bibitem{SH72}
G. Stell and P.C. Hemmer,  J. Chem. Phys.  \textbf{56}, 4274 (1972).

\bibitem{SY76}
M. Silbert and W.H. Young,  Phys. Lett. A  \textbf{58}, 469 (1976).

\bibitem{YA77}
D.A. Young and B. Alder,  Phys. Rev. Lett.  \textbf{38}, 1213 (1977).

\bibitem{LK93}
H. L\"owen and G. Kramposthuber,  Europhys. Lett.  \textbf{23}, 673 (1993).

\bibitem{LALR02}
A.A. Louis, E. Allahyarov, H. L\"owen and R. Roth,  Phys. Rev. E  \textbf{65},
  061407 (2002).

\bibitem{BF94}
P. Bolhuis and D. Frenkel,  Phys. Rev. Lett.  \textbf{72}, 2211 (1994).

\bibitem{BF97}
P. Bolhuis and D. Frenkel,  J. Phys.: Condens. Matter  \textbf{9}, 381 (1997).

\bibitem{RVMN97}
C. Rasc\'on, E. Velasco, L. Mederos and G. Navascu\'es,  J. Chem. Phys.
  \textbf{106}, 6689 (1997).

\bibitem{M98}
L. Mederos,  J. Mol. Liq.  \textbf{76}, 139 (1998).

\bibitem{GGJB99}
A. Galindo, A. Gil-Villegas, G. Jackson and N. Burgess,  J. Phys. Chem. B
  \textbf{103}, 10272 (1999).

\bibitem{LKLLW99}
A. Lang, G. Kahl, C.N. Likos, H. L{\"o}wen and M. Watzlawek,  J. Phys.:
  Condens. Matter  \textbf{11}, {10}{143} (1999).

\bibitem{ZK01}
P. Ziherl and R.D. Kamien,  J. Phys. Chem. B  \textbf{105}, 10147 (2001).

\bibitem{MP03}
G. Malescio and G. Pellicane,  Nat. Mater.  \textbf{2}, 97 (2003).

\bibitem{RS03}
V.N. Ryzhov and S.M. Stishov,  Phys. Rev. E  \textbf{67}, 010201 (2003).

\bibitem{PK08}
G.J. Pauschenwein and G. Kahl,  J. Chem. Phys.  \textbf{129}, 174107 (2008).

\bibitem{BMAGPSSX09}
S.B. Buldyrev, G. Malescio, C. Angell, N. Giovanbattista, S. Prestipino, F.
  Saija, H.E. Stanley and L. Xu,  J. Phys.: Condens. Matter  \textbf{21},
  504106 (2009).

\bibitem{BSB09}
N.M. {Barraz Jr.}, E. Salcedo and M.C. Barbosa,  J. Chem. Phys.  \textbf{131},
  094504 (2009).

\bibitem{ZS09}
S. Zhou and J.R. Solana,  J. Chem. Phys.  \textbf{131}, {204}{503} (2009).

\bibitem{FK10}
J. Fornleitner and G. Kahl,  J. Phys.: Condens. Matter  \textbf{22}, 104118
  (2010).

\bibitem{GSC10}
I. Guill\'en-Escamilla, E. Sch\"oll-Paschinger and R. Casta{\~n}eda-Priego,
  Mol. Phys.  \textbf{108}, 141 (2010).

\bibitem{BL10}
M.N. Bannerman and L. Lue,  J. Chem. Phys.  \textbf{133}, 124506 (2010).

\bibitem{YSH11}
S.B. Yuste, A. Santos and M. {L\'opez de Haro},  Mol. Phys.  \textbf{109}, 987
  (2011).

\bibitem{HHT13}
S.P. Hlushak, P.A. Hlushak and A. Trokhymchuk,  J. Chem. Phys.  \textbf{138},
  164107 (2013).

\bibitem{KH13}
M. Khanpour and R. Hashim,  Phys. Chem. Liq.  \textbf{51}, 203 (2013).

\bibitem{HYS08}
M. {L\'opez de Haro}, S.B. Yuste and A. Santos, {Alternative Approaches to the
  Equilibrium Properties of Hard-Sphere Liquids}, in \emph{{Theory and
  Simulation of Hard-Sphere Fluids and Related Systems}}, edited by A. Mulero,
  \emph{Lecture Notes in Physics}, Vol. 753  (Springer-Verlag, Berlin, 2008),
  pp. 183--245.

\bibitem{S16}
A. Santos, \emph{{A Concise Course on the Theory of Classical Liquids. Basics
  and Selected Topics}}, \emph{Lecture Notes in Physics},  in press
  (Springer-Verlag, Berlin, 2016).

\bibitem{TL93}
Y. Tang and B.C.-Y. Lu,  J. Chem. Phys.  \textbf{99}, 9828 (1993).

\bibitem{TL94a}
Y. Tang and B.C.-Y. Lu,  J. Chem. Phys.  \textbf{100}, 3079 (1994).

\bibitem{TL94b}
Y. Tang and B.C.-Y. Lu,  J. Chem. Phys.  \textbf{100}, 6665 (1994).

\bibitem{T03}
Y. Tang,  J. Chem. Phys.  \textbf{118}, 4140 (2003).

\bibitem{TL97}
Y. Tang and B.C.-Y. Lu,  AIChE J.  \textbf{43}, 2215 (1997).

\bibitem{W63}
M.S. Wertheim,  Phys. Rev. Lett.  \textbf{10}, 321 (1963).

\bibitem{W64}
M.S. Wertheim,  J. Math. Phys.  \textbf{5}, 643 (1964).

\bibitem{T63}
E. Thiele,  J. Chem. Phys.  \textbf{39}, 474 (1963).

\bibitem{YS91}
S.B. Yuste and A. Santos,  Phys. Rev. A  \textbf{43}, 5418 (1991).

\bibitem{LNP_book_note_13_10}
A. Santos  2013, ``Radial Distribution Function for Hard Spheres'', Wolfram
  Demonstrations Project,
  \url{http://demonstrations.wolfram.com/RadialDistributionFunctionForHardSpheres/}.

\bibitem{LSYS05}
J. Largo, J.R. Solana, S.B. Yuste and A. Santos,  J. Chem. Phys.  \textbf{122},
  {084}{510} (2005).

\bibitem{AW92}
J. Abate and W. Whitt,  Queueing Syst.  \textbf{10}, 5 (1992).

\bibitem{private}
S.P. Hlushak, private communication.

\end{thebibliography}

\end{document}